%% file: main.tex
\documentclass{IEEEtran}
\IEEEoverridecommandlockouts
\usepackage{cite}
\usepackage{amsmath,amssymb,amsfonts}
\usepackage{algorithmic}
\usepackage{graphicx}
\usepackage{svg}
\usepackage{pdfpages}
\usepackage{textcomp}
\usepackage{xcolor}
\usepackage{tikz}
\usepackage{array}
\usepackage{float}
\usepackage{pgfplots}
\usepackage{multicol}
\usepackage{adjustbox}
\usepackage{bm}
\usepackage[normalem]{ulem}
\pgfplotsset{compat=1.8}
\def\BibTeX{{\rm B\kern-.05em{\sc i\kern-.025em b}\kern-.08em
    T\kern-.1667em\lower.7ex\hbox{E}\kern-.125emX}}
\begin{document}

\title{Radio Interferometric Calibration with the Exponential Map}

\author{Luca Cabral, Mariano Fernandez-Corazza, Guillermo Gancio, Paula Benaglia
    \thanks{
        This work was supported by UNLP EI003. Corresponding author: \textit{Luca Cabral}.
    }
    \thanks{
        L. Cabral is with the Argentine Institute of Radio astronomy (IAR), UNLP - CONICET - CICPBA, Berazategui, Argentina, and the LEICI Institute of Research in Electronics, Control and Signal Processing, UNLP - CONICET, La Plata, Argentina (e-mail: lcabral@iar.unlp.edu.ar).
    }
    \thanks{
        M. Fernandez-Corazza is with the LEICI Institute of Research in Electronics, Control and Signal Processing, UNLP - CONICET, La Plata, Argentina (e-mail: marianofco@gmail.com).
    }
    \thanks{
        G. Gancio is with the Argentine Institute of Radio astronomy (IAR), UNLP - CONICET - CICPBA, Berazategui, Argentina (e-mail: ganciogm@gmail.com).
    }
    \thanks{
        P. Benaglia is with the Argentine Institute of Radio astronomy (IAR), UNLP - CONICET - CICPBA, Berazategui, Argentina (e-mail: pben.radio@gmail.com).
    }
}

\maketitle

\begin{abstract}
The antenna-elements that make up a radio interferometer form a spatial filter that samples components 
of the Fourier transform of a target radio astronomical source brightness. Along the signal path, there 
are multiplicative and additive perturbation effects that alter the signal and that should be corrected. 
The process of mitigating these perturbation effects is called calibration. In this work, 
we develop a new and fast Maximum Likelihood based estimator of these perturbation effects using the 
Exponential Map and Lie groups. To evaluate its perfomance, we compare it to the Cramér-Rao Lower Bound 
and, to test the estimation time, we compare it with the Expectation Maximization algorithm, another 
fast Maximum Likelihood estimator. Finally, we apply our estimator to a real observation of 
the protoplanetary disk AS 209 made with the Submillimeter Array. We found that our proposed 
estimator meets the Cramér-Rao Lower Bound and it was approximately 40 times faster than the 
Expectation Maximization algorithm.

\end{abstract}


\section{Introduction}
\input{Introduction.tex}

\section{Methods}
\input{Methods.tex}

\section{Simulation and Results}
\input{Simulations.tex}

\section{Conclusions}
We designed a new fast MLE using the Exponential Map to correct direction dependent and 
independent effects in radio interferometric data. The main result was when comparing it with the standard 
EM algorithm, also a fast MLE, where we saw that our estimator is approximately $40$ times 
faster than the EM. The key advantage is the use of Lie groups that simplify the multidimensional 
parametric optimization to one dimensional optimizations. As a secondary contribution to the field, we 
provide the implementation 
of the EM algorithm for the same problem but using the snapshot frequency model instead of using Jones 
matrices.

\begin{IEEEbiography}[{\includegraphics[width=1in,height=1.25in,clip,keepaspectratio]{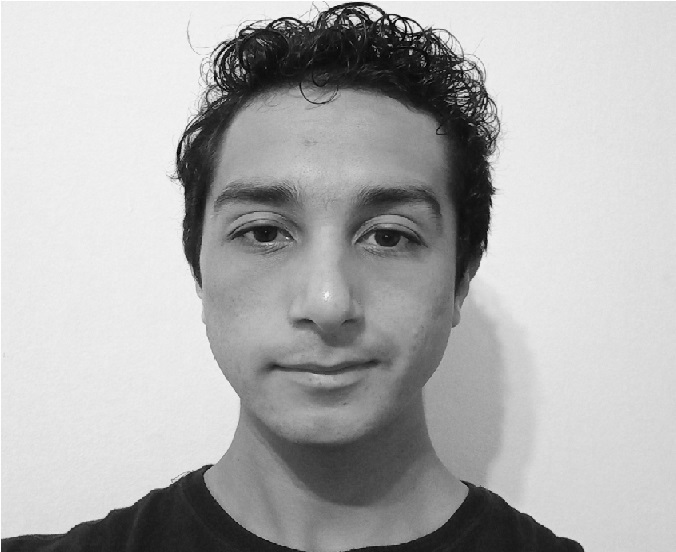}}]{Luca Cabral}
received the Engineering degree from the National University of La Plata (UNLP) in 2021. He is currently a Ph.D. 
student with the Argentine Institute of Radio Astronomy (IAR) and the Instituto LEICI. His research interests are 
in the area of statistical and array signal processing techniques with applications to radio astronomy.
\end{IEEEbiography}
\begin{IEEEbiography}[{\includegraphics[width=1in,height=1.25in,clip,keepaspectratio]{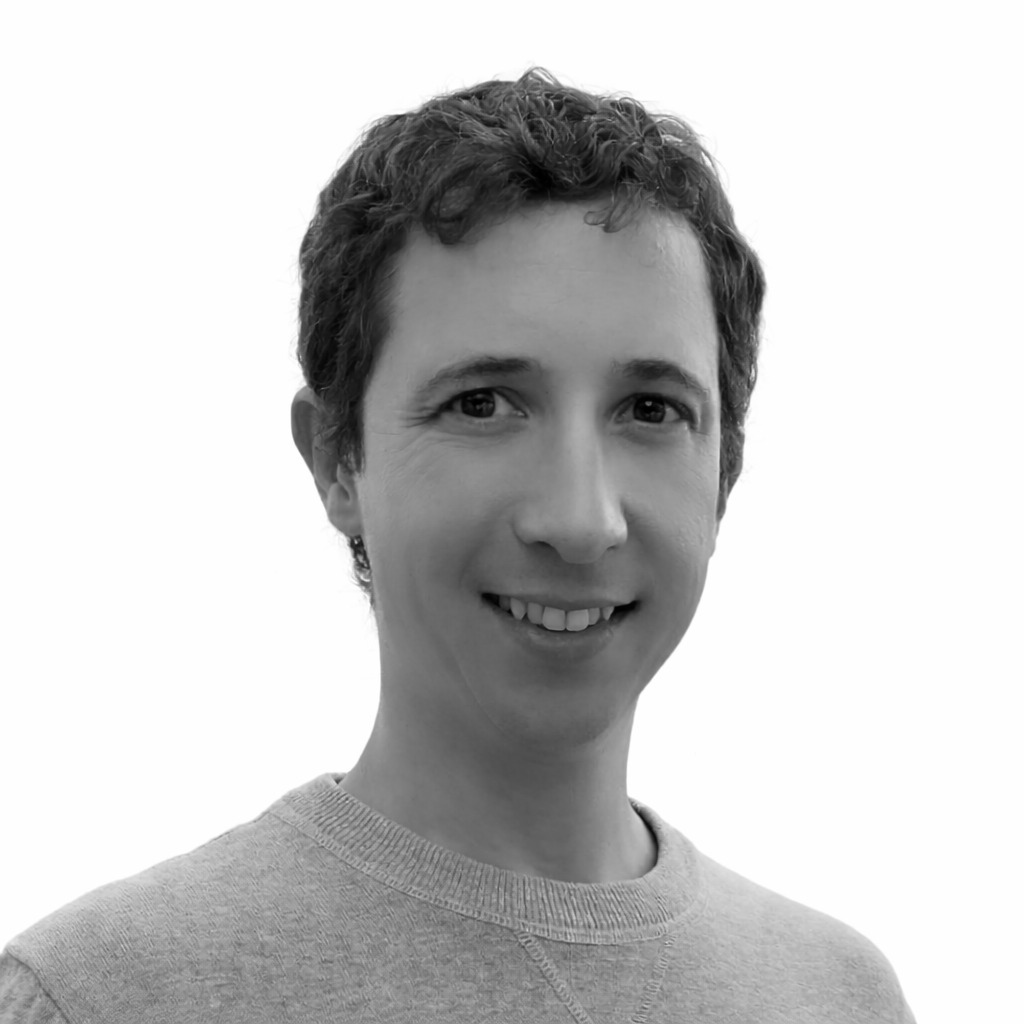}}]{Mariano Fernández Corazza}
received the Ph.D. degree in Engineering from the National University of La Plata 
(UNLP), in 2015. He is currently a Researcher with the Instituto LEICI, UNLP-CONICET, La Plata, Argentina, where 
he leads research in biomedical electromagnetic modeling and signal processing. He has held multiple visiting 
research appointments at the University of Oregon, US and the University of Seville, Spain. His research 
interests include brain electrical and magnetic stimulation, electroencephalography, and electrical impedance 
tomography. He has published over 60 peer-reviewed journal articles and conference papers in these areas and 
has served as Principal Investigator on projects funded by national science agencies.
\end{IEEEbiography}
\begin{IEEEbiography}[{\includegraphics[width=1in,height=1.25in,clip,keepaspectratio]{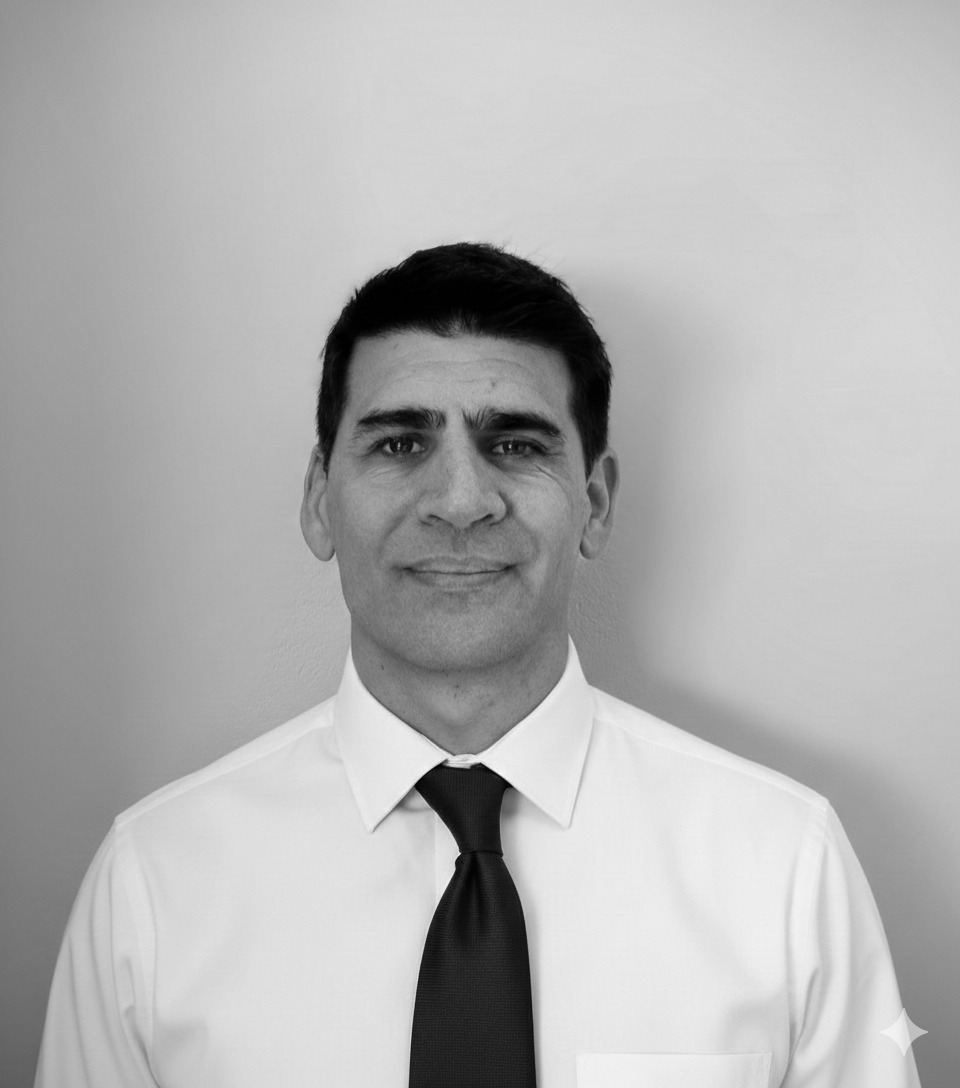}}]{Guillermo Matias Gancio González} 
is Head of Observatory and Project Manager at the Argentine Institute of Radio astronomy (IAR), 
CONICET/CIC/UNLP. With 20+ years in electronic instrumentation for radio astronomy and space, he designs RF 
front ends and digital back ends. He leads LARA, a lunar-orbit microsatellite payload, and MIA-PATHFINDER, a 
16-antenna interferometric array. He developed deep-space digital back ends for CONAE-DSA3 and CLTC-Neuquén, 
contributed to LLAMA, and PuMA pulsar monitoring. He supported Aquarius/SAC-D MWR subsystems and VEX flight 
harnesses. Trained at NASA JPL and NOVA, he co-authors papers in Astronomy \& Astrophysics and MNRAS regularly 
widely cited.
\end{IEEEbiography}
\begin{IEEEbiography}[{\includegraphics[width=1in,height=1.25in,clip,keepaspectratio]{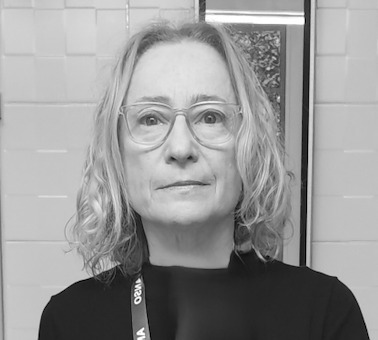}}]{Paula Benaglia}
is Principal Researcher of CONICET at the Argentine Institute of Radio astronomy. Her research focuses primarily 
on detecting and modelling radiation processes of stellar objects in general, with evidence of non-thermal 
radio emission. She obtained her Ph.D. in Astronomy from La Plata National University and a post doctorate at 
the University of São Paulo, Brazil. She was Junior Researcher at NRAO, and Visiting Researcher at ATCA, 
Caltech, NCRA and IIST. She delivered conferences at institutions and scientific meetings in Australia, Spain, 
India, USA, UAE, South Africa. She has been full-time professor at UNLP, invited professor at universities in 
Argentina and abroad, and advisor of many theses. With over 30 years of field research experience, she has 
published 100 peer-reviewed articles on a variety of subjects, with over 2,500 citations, and h-index of 28. 
She was honoured with the Innovative Woman Prize in 2012 and the 2023 Sérsic Prize. 
\end{IEEEbiography}

\end{document}

%% file: Introduction.tex
A radio interferometer is an array of antenna-elements whose baselines, i.e. difference vectors of 
element positions, projected onto the direction of a given astronomical target, sample components of 
the Fourier transform (named $uv$ plane) of the target's brightness, named Visibility function 
\cite{thompson2017interferometry}. Due to 
the geographical distribution of the elements, the electromagnetic interferences at the location of 
the array and the perturbation in the propagation media that affect the signal, 
estimation algorithms are needed to calibrate the spurious effects on the raw data. These errors are 
classified as Direction 
Independent Effects (DIE) and Direction Dependent Effects (DDE). The typical calibration aproach starts 
by assumming that the data is affected by 
unknown deterministic perturbation effects and by additive Gaussian noise. With these assumptions, the 
estimators are designed as Maximum Likelihood Estimators (MLEs), which are known to be asymptotically 
optimal, meaning that their variance meet the Cramér-Rao Bound (CRB) \cite{Kay, van2002optimum}. 
The MLEs involve an 
optimization step, where several algorithms were implemented such as Weighted Least Squares, 
Complex Least Squares, Expectation Maximization (EM) or its improved version, Space Alternating Generalized 
Expectation Maximization (SAGE) \cite{Tol2026, smirnov2015, SAGE, SPT2013, Tol2007, Sarod,RIC_Riemannian}.

The Argentine Institute of Radio astronomy (IAR) is working in a project to deploy the first radio 
interferometer in Argentina, which will be called the Multipurpose Interferometer Array 
(MIA)\cite{Gancio2024,Benaglia2024}. It is called Multipurpose for its broad scientific objectives, which 
include research on fast radio bursts, magnetar flares, supernova remnants, cosmological neutral hydrogen 
(HI) line observations and more. This interferometer will be located at a region of the Andes foothills 
and it will consist of 64 elements of 5-meter diameter dishes. 
The antenna mounts will be alt-azimuth and their feeds will be Vivaldi printed on PCB. The operating 
frequency of the array will be between 1 and 2.3 GHz, with a final bandwith of 1 GHz and 1 arcsec 
resolution at a center frequency of 1.4 GHz. Currently, the IAR is working with a prototype as a proof of 
concept called the ``MIA-Pathfinder'', consisting of three elements of 5-meter diameter dishes. Antenna 
mounts, tracking, motion mechanic systems and radio frequency reception chains are being tested. A 
pathfinder interference test is intended to be carried out towards the end of 2027, when the 
elements of the first stage will be ready. Because of the geographical spread of the elements, the field 
of view when observing and the amount of data when processing an observation, it is necessary to develop 
fast estimators to correct the perturbations on the data.

In this work, we first define the signal model and then introduce an estimator of the perturbations 
based on Lie groups for the DIE case. We then extend this estimator to the full case, 
where DIE and DDE are present. 
To test the estimator performance we compute the CRB for the DIE only case, and 
then we compare it with our estimator's variance error. We compare our estimator with the EM 
algorithm, another fast ML 
estimator, in scenarios with different SNRs, to test which estimator is faster and if its performance 
is the same when the SNR decreases. Finally, as an 
application example, we use our estimator to calibrate the data of an observation of the protoplanetary 
disk AS 209, using a real observation made with the Submillimeter Array (SMA) located in Mauna Kea, 
Hawaii \cite{SMA}.

%% file: Methods.tex
To propose an estimator for DIE and DDE perturbations, first we must define a signal model for the 
ideal case. Then, we add the perturbation effects that affect the observations. With the knowledge of 
the interferometer signal model, we define the Lie groups and then an estimator using them. To evaluate 
the performance of the estimators, we present the CRB and show its expression for the DIE case. 
To compare the computational time, we introduce the EM and its improved version, the SAGE algorithm.

\subsection{Signal model in radio interferometry}

A radio interferomter consists of a set of $L$ antennas (elements) with position vectors $\mathbf{r}_i$ 
with respect to one element considered the reference, 
where $i$ ranges from $0$ to $L-1$. All the elements point to the same direction in the sky. The 
incoming signal from $q$ astronomical sources in the field of view is defined in a vector 
$\mathbf{s}[k]$ of dimension $q \times 1$, where $k$ is the sample number. The direction of the 
sources in the sky are noted by the 
vector $\mathbf{v}_j$, where $j$ ranges from $0$ to $q-1$. The delays related to the signal path from 
source $j$ to element $i$ are stacked in a matrix $\mathbf{A}$ with dimension $L \times q$, where the 
element $i,j$ is $\mathbf{A}_{ij} = e^{-\mathbf{r}_i^T\mathbf{v}_j}$ ($T$ denotes transpose)
\cite{SPT2013}. We now 
define the output signal $\mathbf{x}[k]$ as:
\begin{equation}
    \mathbf{x}[k] = \mathbf{As}[k],
\end{equation}
where each element observes the same sources with different phases because of their geographical 
locations. The output 
of the interferometer is computed as the correlation of the output signal:
\begin{equation}
    E\{\mathbf{x}[k]\mathbf{x}^H[k]\} = \mathbf{A\Sigma A}^H,
\end{equation}
where we consider $\mathbf{s}[k]$ as Wide Sense Stationary \cite{SAGE}, normally distributed with zero 
mean and covariance matrix $\mathbf{\Sigma}$ ($\mathbf{s}[k] \sim \mathcal{N}(\mathbf{0}, 
\mathbf{\Sigma})$), and the superscript $H$ denotes Hermitian transpose. From this expression, it can be 
observed that the interferometer measures the Fourier transform of the brightness distribution of the 
sources in 
the field of view, called the visibilities \cite{thompson2017interferometry}.

\begin{figure}[h!]
    \centering
    \input{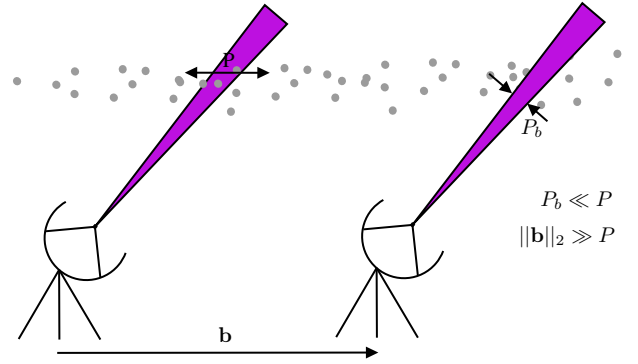}
    \caption{Interference model in the interferometer for ionospheric effects related to the size of the 
    primary beams ($P_b$) and baseline magnitude ($||\mathbf{b}||_2$). $P$ is the section of the sky 
    (ionosphere) where the effects over the incoming signal are assumed to be constant.}
    \label{fig:ionosphere}
\end{figure}

From the astronomical sources to the elements of the radio interferometer array, there are various 
perturbation sources that affect the signal of interest. The phase of the signal is affected by the 
ionosphere and by the 
path-length difference between the elements and the data processing center. The dispersion of the 
primary beam of 
the elements and the gain of the electronics affect the amplitude of the 
signal. Finally, the reception chain adds uncorrelated thermal noise $\mathbf{n}[k]$ (with dimensions 
$L \times 
1$) distributed as $\mathbf{n}[k] \sim \mathcal{N}(\mathbf{0}, \mathbf{\Sigma}_n)$. We assume that the 
effects of the ionosphere are approximately constant over each primary beam-width ($P_b$) but different 
for every element (Fig. \ref{fig:ionosphere}). Then, we can define two diagonal complex matrices where 
all the effects are concentrated in each element of the matrices \cite{Tol2007, Wijnholds2009}. The first 
matrix 
models the DIE ($\mathbf{G}$) and the second models the 
DDE ($\mathbf{D}$). The first matrix is then defined as $\mathbf{G}_{ii} = \gamma_ie^{j\phi_i}$ and 
the second as $\mathbf{D}_{ii} = \beta_ie^{j\varphi_i}$. The output signal becomes:
\begin{equation}
    \mathbf{x}[k] = \mathbf{GADs}[k] + \mathbf{n}[k],
    \label{eq:signal_model}
\end{equation}
and the output of the interferometer is:
\begin{equation}
    \mathbf{R} = E\{\mathbf{x}[k]\mathbf{x}^H[k]\} = \mathbf{GAD\Sigma}\mathbf{D}^H
    \mathbf{A}^H\mathbf{G}^H 
    + \mathbf{\Sigma}_n.
\end{equation}
This last expression is known as the Radio Interferometer Measurement Equation (RIME) 
\cite{thompson2017interferometry,smirnov2011}. Then, the output signal is distributed as 
$\mathbf{x}[k] \sim \mathcal{N}(\mathbf{0}, \mathbf{R}) $.
\subsection{Cramér-Rao Lower Bound}
For a given random vector $\mathbf{x}[n]$ of dimension $r \times 1$, $n = 0,\dots , N-1$ and the 
parameters of interest in a given model stacked into a vector $\mathbf{\theta}^T = 
[\theta_1, \theta_2, \dots, \theta_k]$, the probability density function of the data is denoted as:
\begin{equation}
    f(\mathbf{x}[0], \mathbf{x}[1], \dots, \mathbf{x}[N-1]; \bm{\theta}) = f(\mathbf{x}; 
    \bm{\theta}),
\end{equation}
\noindent where $\bm{\theta}$ parameterizes the distribution. For unbiased 
estimators, i.e. estimators $\hat{\bm{\theta}}$ that satisfy 
$E\{\hat{\bm{\theta}}\} = \bm{\theta}$, the variance represents the quality of the estimator. 
Due to this 
reason, we are interested in the minimum variance an unbiased estimator could reach, 
which is 
given by the Cramér-Rao Lower Bound 
\cite{Tol2007,Wijnholds2008}. If the following regularity condition is satisfied:
\begin{equation}
    E\left\{\dfrac{\partial \ln(f(\mathbf{x}; \bm{\theta}))}{\partial \bm{\theta}}\right\} = 
    \mathbf{0},
\end{equation}
the covariance matrix of any unbiased estimator 
$\mathbf{C}_{\hat{\bm{\theta}}\hat{\bm{\theta}}}$ 
will meet the condition:
\begin{equation}
    \mathbf{C}_{\hat{\bm{\theta}}\hat{\bm{\theta}}} - \dfrac{1}{N}
    \mathbf{F}^{-1}(\bm{\theta}) \geq \mathbf{0},
\end{equation}
\noindent where the $N$ samples of $\mathbf{x}[n]$ are assumed to be independent, the inequality 
$\geq \mathbf{0}$ indicates that the left side is semi-positive definite, and 
$\mathbf{F}(\bm{\theta})$ 
is the Fisher Information Matrix, computed as \cite{Kay}:
\begin{equation}
    \mathbf{F}_{ij}(\bm{\theta}) = -E\left\{\dfrac{\partial^2 \ln(f(\mathbf{x}; \mathbf{\theta}))}
    {\partial \mathbf{\theta}_i\partial \mathbf{\theta}_j}\right\}.
\end{equation}
In the radio interferometric calibration problem, the parameter vector for the DIE only case is:
\begin{equation}
    \bm{\theta} = \begin{pmatrix}
        \bm{\gamma} & \bm{\phi}
    \end{pmatrix}^T,
\end{equation}
and for the full case (DIE + DDE) is:
\begin{equation}
    \bm{\theta} = \begin{pmatrix}
        \bm{\gamma} & \bm{\beta} & \bm{\phi} & \bm{\varphi}
    \end{pmatrix}^T.
\end{equation}
In both cases, the first element of each subparameter vector is removed, because those are related to 
the first antenna, which is taken as the reference. The signal is assumed to be 
Gaussian \cite{Wijnholds2008, SPT2013, kim2024bayesian} and thus, the Fisher Information Matrix can be 
computed using the Slepian-Bangs formula 
\cite{Wijnholds2008,LucaCRB}:
\begin{equation}
    \mathbf{F}(\bm{\theta}) = \left(\dfrac{\partial \text{vec}\left(\mathbf{R}\right)}{\partial 
    \bm{\theta}}\right)^H
    \left( \left(\mathbf{R}^{-1}\right)^* \otimes \mathbf{R}^{-1} \right)
    \left(\dfrac{\partial \text{vec}\left(\mathbf{R}\right)}{\partial \bm{\theta}}\right),
\end{equation}
where:
\begin{equation}
\begin{aligned}
    \dfrac{\partial \text{vec}\left(\mathbf{R}\right)}{\partial \bm{\theta}} = \begin{pmatrix}
        \dfrac{\partial \text{vec}\left(\mathbf{R}\right)}{\partial \bm{\gamma}} &
        \dfrac{\partial \text{vec}\left(\mathbf{R}\right)}{\partial \bm{\phi}}
    \end{pmatrix},
\end{aligned}
\end{equation}
\noindent and the derivatives of $\text{vec}(\mathbf{R})$ with respect to $\bm{\gamma}$ and $\bm{\phi}$ 
for the DIE only case
are \cite{Wijnholds2008}:
\begin{equation}
    \begin{aligned}
    \dfrac{\partial \text{vec}\left(\mathbf{R}\right)}{\partial \bm{\gamma}} = \left(\left( 
        \mathbf{GA\Sigma}\mathbf{A}^H 
    \right)^*\mathbf{\Phi} \circ \mathbf{I} + \mathbf{I} \circ \mathbf{GA\Sigma}\mathbf{A}^H
    \mathbf{\Phi}^*\right)\mathbf{I}_s,
    \end{aligned}
\end{equation}
\begin{equation}
    \dfrac{\partial \text{vec}\left(\mathbf{R}\right)}{\partial \bm{\phi}} = \left(\left( \mathbf{GA\Sigma}\mathbf{A}^H 
    \right)^*\mathbf{G} \circ \mathbf{I} - \mathbf{I} \circ \mathbf{GA\Sigma}\mathbf{A}^H
    \mathbf{G}^*\right)j\mathbf{I}_s.
\end{equation}
$\mathbf{I}_s$ is the selection matrix, which is the identity matrix with the first column removed, 
because the first component of $\mathbf{G}$, i.e. the first antenna, is the reference. 
Its dimension is therefore $L^2 \times L^2-1$. 
The matrix $\mathbf{\Phi}$ is equal to $\mathbf{G}$ without the amplitude components, 
$\mathbf{\Phi}_{ii} = e^{j\phi_i}$. The symbol $\otimes$ denotes the Kronecker product and $\circ$ 
denotes the column-wise Khatri-Rao product \cite{Moon}. 
\subsection{Maximum Likelihood Estimator}
The Maximum Likelihood Estimator is an asymptotically optimal estimator, meaning that it gets closer 
to the CRB as the number of samples $N$ increases. The estimator derives from maximizing the likelihood 
function $\ln(f(\mathbf{x}, \bm{\theta}))$ with respect to $\bm{\theta}$. When we consider 
$\mathbf{s}[k]$ only as an observation of the source sample function, the MLE 
for the full case is derived from minimizing \cite{Kay}:
\begin{equation}
    L(\mathbf{x}, \bm{\theta}) = \sum_{k = 0}^{N-1}||\mathbf{x}[k] - \mathbf{GADs}[k]||_2^2.
    \label{eq:Norma}
\end{equation}
Note that the MLE of Eq. (\ref{eq:Norma}) is to minimize 
the Frobenius Norm of the difference between the data and the signal model.
\subsection{Lie group-based estimator for the DIE case}
\label{subsection:DIE}
We first derive the MLE when DIE only perturbations are present, and we then extend it to include 
the DDE perturbations. The objective is to minimize:
\begin{equation}
    L(\mathbf{x}, \bm{\theta}) = \sum_{k = 0}^{N-1}||\mathbf{x}[k] - \mathbf{GAs}[k]||_2^2,
    \label{eq:NormaDIE}
\end{equation}
with respect to $\bm{\theta}$. The set of all diagonal matrices is not Euclidean but it is a 
manifold, so it has the property of being 
locally Euclidean, allowing to define a tangent space at each point of the manifold 
\cite{SmoothManifolds,Roncagliolo}. The tanget space $T_pM$ is a vector space defined as the set of 
all derivatives 
of the smooth real-valued functions from the manifold $M$ to $\mathbb{R}$ at point $p$, $p \in M$. 
$T_pM$ can be interpreted as the vector space of the best linear approximations of the smooth 
real-valued functions at $p$.
The set of all diagonal matrices defines a group with the usual matrix multiplication, and thus, it
defines a Lie group because it is a group where the set is a manifold. Since we have a Lie group, 
it is parallelizable, i.e it admits a smooth global basis from its tangent space. The basis has $2L$ 
components where the first half of the components are defined as $\mathbf{X}_i = \mathbf{E}_{ii}$, a 
unit element matrix of 
dimension $L \times L$ where all elements are zero except for the element on column and row $i$, which 
is set to $1$ \cite{Moon}. 
The second half of the components are defined in the same way but multiplied by the imaginary unit, 
$\mathbf{X}_{i+L} = j\mathbf{E}_{ii}$, with $i$ ranging from $0$ to $L-1$. Let $\mathbf{X}_a, \mathbf{X}_b$ 
be elements of the tangent space, a 
metric can be defined as $<\mathbf{X}_a, \mathbf{X}_b> =  \text{tr}\{\mathbf{X}_a^H\mathbf{X}_b\} = 
||\mathbf{X}_a^H\mathbf{X}_b||_F$, where $\text{tr}$ denotes the trace operator and ``F'' means the 
Frobenius norm.

\begin{figure}[h!]
    \centering
    \input{Figures/ExpMap.tex}
    \caption{Exponential Map generated by the vector $\mathbf{X}$, which is an element of the tangent 
    space of the manifold $M$ at point $\mathbf{I}_0$. The starting point of the curve is 
    $\mathbf{I}_0$.}
    \label{fig:ExpMap}
\end{figure}
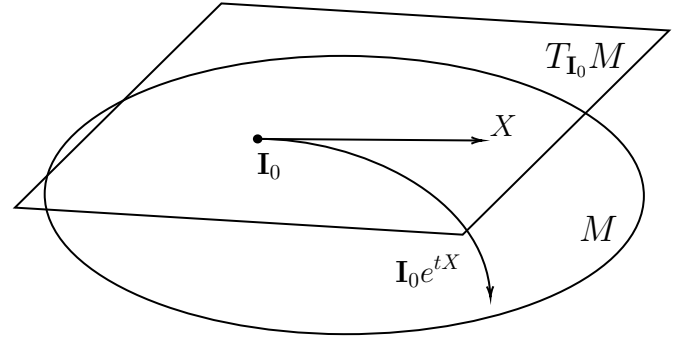

An important characteristic of the Lie groups is the Exponential Map and the one-parameter subgroup. 
The Exponential Map is a map from the tangent space to the manifold and is computed with the 
exponential matrix $\mathbf{I}_0e^{t\mathbf{X}}$ (Fig. \ref{fig:ExpMap}), where $\mathbf{I}_0 \in M$ 
is the 
starting point, $\mathbf{X} \in T_{\mathbf{I_0}}M$ is the initial velocity and $t \in \mathbb{R}$ is 
the parameter which for $t = 0$ gives the starting point. The curve generated by the 
exponential map is called the one-parameter subgroup.
The one-parameter subgroup gives an easy way to search for the diagonal matrix that minimizes Eq. 
(\ref{eq:NormaDIE}). Defining $\mathbf{G} = \mathbf{I}_0e^{t\mathbf{X}}$, where $\mathbf{X}$ is a linear 
combination of the basis $\mathbf{X}_i$:
\begin{equation}
    \mathbf{X} = \sum_{i = 0}^{2L-1}\omega_i \mathbf{X}_i,
\end{equation}
with $\omega_i \in \mathbb{C}$ and $\mathbf{I}_0$ is the starting point of the curve, replacing 
it in Eq. (\ref{eq:NormaDIE}), the functional becomes:
\begin{equation}
    L(\mathbf{x}, \bm{\theta}) = \sum_{k = 0}^{N-1}||\mathbf{x}[k] - 
    \mathbf{I}_0e^{t\mathbf{X}}\mathbf{As}[k]||_2^2.
    \label{eq:NormaDIE2}
\end{equation}
The goal is finding a vector pointing in the direction that minimizes Eq. (\ref{eq:NormaDIE2}). 
It can be obtained by first evaluating the functional in a point slightly off from the starting point 
\cite{SmoothManifolds, smith2005covariance}, 
which allows us to approximate the exponential term as a linear polinomial:
\begin{equation}
    \begin{aligned}
        \mathbf{I}_0e^{t\mathbf{X}} = \mathbf{I}_0\sum_{z = 0}^{\infty}\dfrac{1}{z!}(t\mathbf{X})^z \approx
    \mathbf{I}_0\left(\mathbf{I} + t\mathbf{X}\right) \\ = \mathbf{I}_0\left(\mathbf{I} + 
    \sum_{i=0}^{2L-1}h_i\mathbf{X}_i\right),
    \end{aligned}
\end{equation}
where $h_i = t\omega_i$. Then, we can consider our 
functional as a function of this perturbation as follows:
\begin{equation}
    \begin{aligned}
    L\left( \mathbf{I}_0\left( \mathbf{I} + \sum_{i = 0}^{2L-1}h_i\mathbf{X}_i \right) \right) = 
    L\left( h_0, h_1, \dots, h_{2L-1} \right) = L(\mathbf{h}).
    \end{aligned}
\end{equation}
The derivatives at the starting point are then computed as:
\begin{equation}
    \begin{aligned}
    \nabla L(\mathbf{h}) = \begin{pmatrix}
        \dfrac{\partial L(\mathbf{h})}{\partial h_0} &
        \dfrac{\partial L(\mathbf{h})}{\partial h_1} &
        \cdots &
        \dfrac{\partial L(\mathbf{h})}{\partial h_{2L-1}}
    \end{pmatrix}\Bigg|_{\mathbf{h} = \mathbf{0}} 
        \\ = \begin{pmatrix}
            \alpha_0 & \alpha_1 & \cdots & \alpha_{2L-1}
        \end{pmatrix},
    \end{aligned}
    \label{eq:gradient}
\end{equation}
and therefore the velocity is:
\begin{equation}
    \mathbf{X} = \sum_{i = 0}^{2L-1}\alpha_i\mathbf{X}_i.
    \label{eq:defX}
\end{equation}
\noindent To make the computation of Eq. (\ref{eq:gradient}) easier, we can rewrite the functional of 
Eq. (\ref{eq:NormaDIE2}) using the trace as follows:
\begin{equation}
    \begin{aligned}
        L(\mathbf{x}, \bm{\theta}) = \text{tr}\left\{\sum_{k = 0}^{N-1}(\mathbf{x}[k] - 
        \mathbf{GAs}[k])^H(\mathbf{x}[k] - 
        \mathbf{GAs}[k])\right\}.
    \end{aligned}
    \label{eq:trace1}
\end{equation}
Using the cyclic property of the trace, Eq. (\ref{eq:trace1}) becomes:
\begin{equation}
    \begin{aligned}
        L(\mathbf{x}, \bm{\theta}) = \text{tr}\left\{\mathbf{R}_x - \mathbf{C}_{sx}
        \mathbf{GA} - \right.\\
        \left.\mathbf{C}_{xs}\mathbf{A}^H\mathbf{G}^H + \mathbf{A}\mathbf{C}_{ss}\mathbf{A}^H
        \mathbf{G}^H\mathbf{G} \right\},
    \end{aligned}
    \label{eq:trace}
\end{equation}
where $\mathbf{R}_x$ is the sample covariance matrix of $\mathbf{x}$, $\mathbf{C}_{sx}$ is the sample 
covariance matrix
between the source and the interferometer output, $\mathbf{C}_{xs} = \mathbf{C}_{sx}^H$ and 
$\mathbf{C}_{ss}$ is the sample covariance matrix of the astronomical sources. In this calculation, we 
dropped the constant $N$ which multiplies all the terms. Because we consider the source 
as unknown, we first minimize Eq. (\ref{eq:trace1}) with respect to the source signal, leading to 
the Minimum Variance Distortionless Response (MVDR) \cite{van2002optimum}:
\begin{equation}
    \mathbf{C}_{ss} = \mathbf{W}\mathbf{R}_x\mathbf{W}^H,
\end{equation}
\begin{equation}
    \mathbf{C}_{sx} = \mathbf{W}\mathbf{R}_x,
\end{equation}
\begin{equation}
    \mathbf{W} = \left(\left(\mathbf{GA}\right)^H\mathbf{GA}\right)^{-1}\left(\mathbf{GA}\right)^H.
    \label{eq:MVDR}
\end{equation}
Using the 
properties $\text{tr}\{\mathbf{A}^H\mathbf{B}\} = \text{vec}^H(\mathbf{A})\text{vec}(\mathbf{B})$ and 
$\text{vec}\{\mathbf{ABC}\} = \left( \mathbf{C}^T \otimes \mathbf{A} \right)\text{vec}(\mathbf{B}) $, 
where $\text{vec}$ is the vectorization operator, Eq. (\ref{eq:trace}) can be written as:
\begin{equation}
    \begin{aligned}
        L(\mathbf{x}, \bm{\theta}) = \text{tr}\left\{\mathbf{R}_x\right\} -
        \text{vec}^H\left( \mathbf{C}_{sx}^H \mathbf{A}^H \right)\text{vec}(\mathbf{G}) - \\
        \text{vec}^H\left( \mathbf{A}\mathbf{C}_{xs}^H \right)\text{vec}\left(\mathbf{G}^H\right) + \\
        \text{vec}^H\left(\mathbf{G}\right)\left( \mathbf{A}^* \mathbf{C}_{ss}^*\mathbf{A}^T \otimes 
        \mathbf{I}\right)\text{vec}(\mathbf{G}).
    \end{aligned}
    \label{eq:L21}
\end{equation}
Reminding that $\mathbf{G}$ is diagonal, the derivatives of Eq. (\ref{eq:trace}), with respect to the first 
half components of $\mathbf{h}$, to compute $\mathbf{X}$ result:
\begin{equation}
    \begin{aligned}
        \nabla_1 L(\mathbf{x}, \bm{\theta}) = -
        \text{vec}^H\left( \mathbf{C}_{sx}^H \mathbf{A}^H \right)(\mathbf{G} \circ \mathbf{I}) - \\
        \text{vec}^H\left( \mathbf{A}\mathbf{C}_{xs}^H \right)(\mathbf{G} \circ \mathbf{I})^* + \\
        (\mathbf{G} \circ \mathbf{I})^H\left( \mathbf{A}^* \mathbf{C}_{ss}^*\mathbf{A}^T \otimes 
        \mathbf{I}\right)\text{vec}(\mathbf{G}) + \\
        \text{vec}^H\left(\mathbf{G}\right)\left( \mathbf{A}^* \mathbf{C}_{ss}^*\mathbf{A}^T \otimes 
        \mathbf{I}\right)(\mathbf{G} \circ \mathbf{I}).
    \end{aligned}
    \label{eq:X}
\end{equation}
The full calculation from Eq. (\ref{eq:trace}) to Eq. (\ref{eq:X}) can be found in the 
Supplementary material. Equation (\ref{eq:X}) shows the 
derivates to compute the first half components of $\mathbf{X}$, 
the second half is the same but multiplied by the imaginary unit $j$. With 
$\mathbf{X}$ calculated, the curve to search the matrix to minimize the functional is now defined, 
so the 
next step is to minimize the functional with respect to $t$. To solve this step we use the gradient 
descent algorithm as follows:
\begin{equation}
    t_{k+1} = t_k - g\dfrac{dL(\mathbf{x}, \bm{\theta})}{dt}\Bigg|_{t_k},
\end{equation}
where $k$ is the iteration number and $g$ is the step gain. The derivative with respect to $t$ of 
$e^{t\mathbf{X}}$ is $\mathbf{X}e^{t\mathbf{X}}$, thus:
\begin{equation}
    \begin{aligned}
        \dfrac{dL(\mathbf{x}, \bm{\theta})}{dt} = \text{tr}\left\{
        - \mathbf{A}\mathbf{C}_{sx}\mathbf{I}_0\mathbf{X}e^{t\mathbf{X}}\right.
        - \mathbf{C}_{xs}\mathbf{A}^H\left(\mathbf{I}_0\mathbf{X}e^{t\mathbf{X}}\right)^H \\
        + \mathbf{I}_0\mathbf{X}e^{t\mathbf{X}}\mathbf{A}\mathbf{C}_{ss}\mathbf{A}^H
        \mathbf{G}^H 
        \left. + \mathbf{G}\mathbf{A}\mathbf{C}_{ss}\mathbf{A}^H
        \left(\mathbf{I}_0\mathbf{X}e^{t\mathbf{X}}\right)^H\right\}.
    \end{aligned}
    \label{eq:gradiente}
\end{equation}
Once the algorithm reaches the 
convergence criterion, the estimation of the DIE matrix is:
\begin{equation}
    \hat{\mathbf{G}} = \mathbf{I}_0e^{t_{\text{end}}\mathbf{X}},
\end{equation}
where $\mathbf{X}$ was computed using Eq. (\ref{eq:defX}) and Eq. (\ref{eq:X}), and $t_{\text{end}}$ was 
the last iteration of the 
gradient descent algorithm. The next step is to repeat this procedure replacing $\mathbf{G}$ with 
$\hat{\mathbf{G}}$ in Eqs. (\ref{eq:MVDR}) and (\ref{eq:X}), and $\mathbf{I_0}$ with $\hat{\mathbf{G}}$ as 
starting point in Eq. (\ref{eq:gradiente}) until reaching convergence. For the first 
iteration we choose $\hat{\mathbf{G}} = \mathbf{I}$. The procedure can 
be seen as iterative: moving through curves towards the direction that minimizes Eq. 
(\ref{eq:NormaDIE2}). It should be noted that the advantage of using Lie groups is that they allow 
the conversion of a multidimensional optimization problems into one-dimension problems where 
the optimization variable is a single scalar parameter. Of course, this comes at the expense of more 
but much faster iterations. Overall, this results in faster estimator implementations.
\subsection{Lie group-based estimator for the full case}
Considering both DIE and DDE, the goal is to minimize Eq. (\ref{eq:Norma}). Operating 
similarly to the DIE case, Eq. (\ref{eq:trace}) is modified as follows:
\begin{equation}
    \begin{aligned}
        L(\mathbf{x}, \bm{\theta}) = \text{tr}\left\{\mathbf{R}_x - \mathbf{C}_{sx}
        \mathbf{GAD} - \right.\\
        \left.\mathbf{C}_{xs}\mathbf{D}^H\mathbf{A}^H\mathbf{G}^H + \mathbf{C}_{ss}\mathbf{D}^H
        \mathbf{A}^H\mathbf{G}^H\mathbf{GAD} \right\}.
    \end{aligned}
    \label{eq:traceDDE}
\end{equation}
Operating in Eq. (\ref{eq:traceDDE}) we can group the terms that depend only 
on $\mathbf{G}$ and $\mathbf{D}$:
\begin{equation}
    \begin{aligned}
        L(\mathbf{x}, \bm{\theta}) = \text{tr}\left\{\mathbf{R}_x \right\} - 
        \text{vec}^H\left( \mathbf{A}^H \right)\left( \mathbf{G} \otimes \mathbf{D} \right)
        \text{vec}\left( \mathbf{C}_{sx} \right) - \\
        \text{vec}^H\left( \mathbf{A} \right)\left( \mathbf{G} \otimes \mathbf{D} \right)^H
        \text{vec}\left( \mathbf{C}_{xs} \right) + \\
        \text{vec}^H\left( \mathbf{A}^H \right)\left( \mathbf{G} \otimes \mathbf{D} \right)
        \left( \mathbf{I} \otimes \mathbf{C}_{ss} \right)
        \left( \mathbf{G} \otimes \mathbf{D} \right)^H\text{vec}\left( \mathbf{A}^H \right).
    \end{aligned}
\end{equation}
Defining $\mathbf{Y} = \left( \mathbf{G} \otimes \mathbf{D} \right)$, the Kronecker product of two 
diagonal matrices is another diagonal matrix and thus, we face again the problem of estimating the 
elements of a diagonal matrix. The difference with the DIE only case is that the perturbation 
effects are 
mixed in the new matrix $\mathbf{Y}$. To make the computation of the derivatives easier, we use 
the fact that
$\text{vec}\left(L(\mathbf{x}, \bm{\theta})\right) = L(\mathbf{x}, \bm{\theta})$:
\begin{equation}
    \begin{aligned}
        L(\mathbf{x}, \bm{\theta}) = \text{tr}\left\{\mathbf{R}_x \right\} - 
        \left( \text{vec}^T\left( \mathbf{C}_{sx} \right) \otimes  
        \text{vec}^H\left( \mathbf{A}^H \right)\right)\text{vec}(\mathbf{Y}) - \\
        \left( \text{vec}^T\left( \mathbf{A}^H \right) \otimes  
        \text{vec}^H\left( \mathbf{C}_{xs}^H \right)\right)\text{vec}(\mathbf{Y}^H) + \\
        \text{vec}^H\left( \mathbf{A}^H \right)\mathbf{Y}
        \left( \mathbf{I} \otimes \mathbf{C}_{ss} \right)
        \mathbf{Y}^H\text{vec}\left( \mathbf{A}^H \right).
    \end{aligned}
    \label{eq:DDE}
\end{equation}
The derivatives are then:
\begin{equation}
    \begin{aligned}
        \nabla_1 L(\mathbf{x}, \bm{\theta}) = -\left( \text{vec}^T\left( \mathbf{C}_{sx} \right) 
        \otimes  \text{vec}^H\left( \mathbf{A}^H \right)\right)
        \left( \mathbf{Y} \circ \mathbf{I} \right) - \\
        \left( \text{vec}^T\left( \mathbf{A}^H \right) \otimes  
        \text{vec}^H\left( \mathbf{C}_{xs}^H \right)\right)
        \left( \mathbf{Y} \circ \mathbf{I} \right)^* + \\
        \left(\left(\left(\mathbf{I} \otimes \mathbf{C}_{ss} \right)\mathbf{Y}^H\text{vec}^H
        \left(\mathbf{A}\right)\right)^T \otimes \text{vec}^H\left(\mathbf{A}^H\right)\right)
        \left( \mathbf{Y} \circ \mathbf{I} \right) + \\
        \left(\text{vec}^*\left(\mathbf{A}\right) \otimes \left( \text{vec}^H\left(\mathbf{A}^H\right)
        \mathbf{Y}\left(\mathbf{I} \otimes \mathbf{C}_{ss} \right) \right)^T\right)
        \left( \mathbf{Y} \circ \mathbf{I} \right)^*.
    \end{aligned}
    \label{eq:XDDE}
\end{equation}
Now $\mathbf{X}$ is computed for the full case using Eq. (\ref{eq:defX}). 
The full calculation from Eq. (\ref{eq:traceDDE}) to Eq. (\ref{eq:DDE}) and to get Eq. 
(\ref{eq:XDDE}) can be found in the 
Supplementary material.

To perform the gradient descent step we take the derivative of Eq. (\ref{eq:DDE}) with respect to $t$ 
reminding that $\mathbf{Y} = \mathbf{I}_0e^{t\mathbf{X}}$:
\begin{equation}
    \begin{aligned}
        \dfrac{dL(\mathbf{x}, \bm{\theta})}{dt} = -\left( \text{vec}^T\left( \mathbf{C}_{sx} \right) 
        \otimes  \text{vec}^H\left( \mathbf{A}^H \right)\right)
        \left( \mathbf{Y}' \circ \mathbf{I} \right) - \\
        \left( \text{vec}^T\left( \mathbf{A}^H \right) \otimes  
        \text{vec}^H\left( \mathbf{C}_{xs}^H \right)\right)
        \left( \mathbf{Y}' \circ \mathbf{I} \right)^H + \\
        \left(\left(\left(\mathbf{I} \otimes \mathbf{C}_{ss} \right)\mathbf{Y}^H\text{vec}^H
        \left(\mathbf{A}\right)\right)^T \otimes \text{vec}^H\left(\mathbf{A}^H\right)\right)
        \left( \mathbf{Y}' \circ \mathbf{I} \right) + \\
        \left(\text{vec}^*\left(\mathbf{A}\right) \otimes \left( \text{vec}^H\left(\mathbf{A}^H\right)
        \mathbf{Y}\left(\mathbf{I} \otimes \mathbf{C}_{ss} \right) \right)^T\right)
        \left( \mathbf{Y}' \circ \mathbf{I} \right)^H,
    \end{aligned}
    \label{eq:tDDE}
\end{equation}
where $\mathbf{Y}' = \mathbf{I}_0\mathbf{X}e^{t\mathbf{X}}$. From Eqs. (\ref{eq:XDDE}) and 
(\ref{eq:tDDE}), the procedure for the perturbation estimation is the same as in Section 
\ref{subsection:DIE}. 
An important diffence with the DIE case are the estimations of $\mathbf{C}_{ss}$ and $\mathbf{C}_{sx}$, 
because we now need to compute the product $\mathbf{GAD}$. For this purpose, we use the identity:
\begin{equation}
    \begin{aligned}
    \text{vec}\left( \left(\mathbf{GAD}\right)^H \right) = 
    \text{vec}\left(\mathbf{D}^H\mathbf{A}^H \mathbf{G}^H\right) = \\
    \left( \mathbf{G}^* \otimes \mathbf{D}^H \right)
    \text{vec}(\mathbf{A}^H) = \left( \mathbf{G} \otimes \mathbf{D} \right)^H
    \text{vec}(\mathbf{A}^H) = \\\mathbf{Y}^H\text{vec}(\mathbf{A}^H).
    \end{aligned}
    \label{eq:GADkron}
\end{equation}
In summary, we first compute the product $\mathbf{Y}^H\text{vec}(\mathbf{A}^H)$, then we 
rearrange the 
elements and finally, compute the Hermitian transpose to obtain the product $\mathbf{GAD}$.

\subsection{EM and the SAGE algorithms}
To compare our algorithm, we implemented the classical EM
algorithm for our signal model, referred as the frequency snapshot model\cite{van2002optimum}.
The SAGE algorithm, which is an improved version of the EM algorithm, was previously implemented for 
the calibration problem \cite{SAGE}, 
but the signal model was defined in terms of Jones matrices \cite{smirnov2011}. 

The output of an interferometer has the form of Eq. (\ref{eq:signal_model}). Now suppose that we 
want to observe each source individually:
\begin{equation}
    \mathbf{m}_i[k] = \mathbf{GA}_is_i[k] + \mathbf{n}_i[k],
    \label{eq:m_i}
\end{equation}
where $\mathbf{A}_i$ is the $i$-th column of $\mathbf{A}$ and $\mathbf{n}_i[k]$ is such that:
\begin{equation}
    \mathbf{n}[k] = \sum_{i=0}^{q-1}\mathbf{n}_i[k],
\end{equation}
where $\mathbf{n}_i[k] \sim \mathcal{N}\left(\mathbf{0}, (1/q)\mathbf{\Sigma}_n\right)$. We refer to the 
$\mathbf{m}_i[k]$ components as the incomplete data, and to $\mathbf{x}[k]$ as the 
complete data. The relation between them is:
\begin{equation}
    \begin{aligned}
    \mathbf{x}[k] = \begin{pmatrix}
        \mathbf{I}_q & \mathbf{I}_q & \cdots & \mathbf{I}_q
    \end{pmatrix}\begin{pmatrix}
        \mathbf{m}_0^T[k] & \mathbf{m}_1^T[k] & \cdots & \mathbf{m}_{q-1}^T[k]
    \end{pmatrix}^T \\ = \mathbf{H}\mathbf{m}^T,
    \end{aligned}
\end{equation}
where $\mathbf{H}$ is the linear transformation that combines the incomplete data and is composed of 
$q$ identity matrices, thus $\mathbf{H}$ has dimensions of $q \times q^2$. As the 
$\mathbf{m}_i[k]$ are unknown, we cannot maximize their likelihood function to estimate 
$\mathbf{\theta}$. Instead 
of that, we first take the expectation of $\mathbf{m}_i[k]$ subject to having 
observed $\mathbf{x}[k]$ (Expectation step) and then maximize the likelihood function with respect to 
$\bm{\theta}$ 
(Maximization step). It is important to highlight that because of the decomposition in 
the incomplete data, the EM steps should be computed $q$ times, one per source. The 
complete and incomplete data are jointly Gaussian, so the result of the Expectation 
step is \cite{van2002optimum}:
\begin{equation}
    \hat{\mathbf{m}}_i[k] = \hat{\mathbf{G}}\mathbf{A}_i\hat{s}_i[k] + \dfrac{1}{L}\left(\mathbf{x}[k] - 
    \hat{\mathbf{G}}\mathbf{A}\hat{\mathbf{s}}[k]\right).
    \label{eq:E}
\end{equation}
As in Subsection \ref{subsection:DIE}, $\hat{\mathbf{s}}[k]$ is computed using the MVDR:
\begin{equation}
    \hat{\mathbf{s}}[k] = \mathbf{Wx}[k],
\end{equation}
where $\mathbf{W}$ is computed using Eq. (\ref{eq:MVDR}). The Maximization step consists in minimizing:
\begin{equation}
    L_2(\mathbf{m}, \bm{\theta}) = \sum_{k = 0}^{N-1}\sum_{i = 0}^{q-1}||\mathbf{m}_i[k] - 
    \mathbf{GA}_is_i[k]||_2^2,
    \label{eq:EMmin}
\end{equation}
with respect to the parameters $\mathbf{\theta}$, which are distributed within $\mathbf{G}$ as in the 
case of Subsection \ref{subsection:DIE}. 
This step is solved using the gradient descent algorithm. 
To apply it in this problem, we have to find the derivatives with 
respect to each parameter. Reminding that $\mathbf{G}_{ii} = \gamma_ie^{j\phi_i}$ and 
operating as in previous subsections, the derivatives of the 
likelihood function with respect to the parameters are:
\begin{equation}
    \dfrac{dL_2(\mathbf{m}, \bm{\theta})}{d\bm{\theta}} = \begin{pmatrix}
        \dfrac{dL_2(\mathbf{m}, \bm{\theta})}{d\bm{\gamma}} & 
        \dfrac{dL_2(\mathbf{m}, \bm{\theta})}{d\bm{\phi}}
    \end{pmatrix},
\end{equation}
\begin{equation}
    \begin{aligned}
        \dfrac{dL_2(\mathbf{m}, \bm{\theta})}{d\bm{\gamma}} = \sum_{i = 0}^{q-1} - 
        \text{vec}^H\left(
        \mathbf{S}_{s_im_i}\mathbf{A}_i^H \right)\left( \mathbf{\Phi} \circ \mathbf{I} \right) - \\
        \text{vec}^H\left( \mathbf{A}_i\mathbf{S}_{s_im_i}^H \right)
        \left( \mathbf{\Phi}^* \circ \mathbf{I} \right) + \\
        \left( \mathbf{\Phi} \circ \mathbf{I} \right)^T\left( \mathbf{I} \otimes \mathbf{A}_i
        \mathbf{S}_{s_is_i}\mathbf{A}_i^H \right)\text{vec}\left( \mathbf{G}^H \right) + \\ 
        \text{vec}\left( \mathbf{G} \right)\left( \mathbf{I} \otimes \mathbf{A}\mathbf{S}_{s_is_i}
        \mathbf{A}^H \right)\left( \mathbf{\Phi}^* \circ \mathbf{I} \right),
    \end{aligned}
    \label{eq:dLgamma}
\end{equation}
\begin{equation}
    \begin{aligned}
        \dfrac{dL_2(\mathbf{m}, \bm{\theta})}{d\bm{\phi}} = \sum_{i = 0}^{q-1} - 
        \text{vec}^H\left(
        \mathbf{S}_{s_im_i}\mathbf{A}_i^H \right)\left( j\mathbf{G} \circ \mathbf{I} \right) - \\
        \text{vec}^H\left( \mathbf{A}_i\mathbf{S}_{m_is_i} \right)\left( -j\mathbf{G}^* \circ 
        \mathbf{I} \right) + \\
        \left( j\mathbf{G} \circ \mathbf{I} \right)^T\left( \mathbf{I} \otimes \mathbf{A}_i
        \mathbf{S}_{s_is_i}\mathbf{A}_i^H \right)\text{vec}\left( \mathbf{G}^H \right) + \\ 
        \text{vec}\left( \mathbf{G} \right)\left( \mathbf{I} \otimes \mathbf{A}_i\mathbf{S}_{s_is_i}
        \mathbf{A}_i^H \right)\left( -j\mathbf{G}^* \circ \mathbf{I} \right),
    \end{aligned}
    \label{eq:dLphi}
\end{equation}
where $\mathbf{S}_{m_is_i}$ is the sample covariance matrix of the incomplete data vector 
$i$ and source signal $i$, $\mathbf{S}_{m_is_i} = \mathbf{S}_{s_im_i}^H$ and $\mathbf{S}_{s_is_i}$ is 
the sample covariance matrix of source $i$. The full computation from Eq. (\ref{eq:EMmin}) to 
get Eqs. (\ref{eq:dLgamma}) and (\ref{eq:dLphi}) is 
detailed in the Supplementary material. Knowing the derivatives, we then compute:
\begin{equation}
    \bm{\theta}_{r+1} = \bm{\theta}_{r} - 
    g\dfrac{dL_2(\mathbf{m}, \bm{\theta})}{d\bm{\theta}^T}\Bigg|_{\bm{\theta}_r},
\end{equation}
where $r$ is the iteration number and $g$ is the step gain. Once the gradient descent algorithm reaches 
the convergence 
criterion, the estimation of the DIE matrix $\mathbf{G}$ is computed as:
\begin{equation}
    \bm{\theta}_{\text{end}} = \begin{pmatrix}
        \hat{\gamma}_1 & \hat{\gamma}_2 & \cdots & \hat{\gamma}_{L-1} & \hat{\phi}_1 & \hat{\phi}_2 &
        \cdots & \hat{\phi}_{L-1}
    \end{pmatrix},
\end{equation}
\begin{equation}
    \hat{\mathbf{G}}_{ii} = \hat{\gamma}_ie^{j\hat{\phi}_i},
    \label{eq:mapeo}
\end{equation}
where $\bm{\theta}_{\text{end}}$ is the last iteration of the gradient descent step and $i$ 
ranges from $1$ to $L-1$, because the first antenna is taken as the reference. The next step is to 
repeat the Expectation and Maximization steps (Eqs. (\ref{eq:E}) to (\ref{eq:mapeo})) using the 
last estimation of $\hat{\mathbf{G}}$ until reaching 
convergence. For the first iteration we choose $\hat{\mathbf{G}} = \mathbf{I}$.

The SAGE algorithm is an improved version which modifies the definition of the incomplete data 
gathering near sources into one vector $\mathbf{m}_i[k]$ to reduce the computational cost. It can be 
noticed that when there is only one source in observation, there is no decomposition into individual 
sources as in Eq. (\ref{eq:m_i}), therefore both the EM and SAGE algorithms are the same.

%% file: Figures/ExpMap.tex
\tikzset{every picture/.style={line width=0.75pt}} 

\begin{tikzpicture}[x=0.75pt,y=0.75pt,yscale=-0.5,xscale=0.5]

\draw   (30.21,192.24) .. controls (30.53,115.1) and (165.43,53.07) .. (331.51,53.68) .. controls (497.6,54.29) and (631.97,117.31) .. (631.65,194.44) .. controls (631.32,271.57) and (496.43,333.61) .. (330.34,333) .. controls (164.26,332.39) and (29.89,269.37) .. (30.21,192.24) -- cycle ;
\draw   (449.74,233.14) -- (0.33,205.97) -- (207.59,1.03) -- (657,28.2) -- cycle ;
\draw    (243.97,137.02) .. controls (344.63,136.31) and (472.49,194.68) .. (477.61,294.49) ;
\draw [shift={(477.68,296)}, rotate = 267.78] [color={rgb, 255:red, 0; green, 0; blue, 0 }  ][line width=0.75]    (10.93,-3.29) .. controls (6.95,-1.4) and (3.31,-0.3) .. (0,0) .. controls (3.31,0.3) and (6.95,1.4) .. (10.93,3.29)   ;
\draw  [fill={rgb, 255:red, 0; green, 0; blue, 0 }  ,fill opacity=1 ] (240.2,137.02) .. controls (240.2,135.06) and (241.89,133.47) .. (243.97,133.47) .. controls (246.05,133.47) and (247.73,135.06) .. (247.73,137.02) .. controls (247.73,138.99) and (246.05,140.58) .. (243.97,140.58) .. controls (241.89,140.58) and (240.2,138.99) .. (240.2,137.02) -- cycle ;
\draw    (243.97,137.02) -- (467.45,138.61) ;
\draw [shift={(469.45,138.62)}, rotate = 180.41] [color={rgb, 255:red, 0; green, 0; blue, 0 }  ][line width=0.75]    (10.93,-3.29) .. controls (6.95,-1.4) and (3.31,-0.3) .. (0,0) .. controls (3.31,0.3) and (6.95,1.4) .. (10.93,3.29)   ;

\draw (240.99,148.97) node [anchor=north west][inner sep=0.75pt]  [xscale=0.8,yscale=0.8] [align=left] {{\Large $\displaystyle \mathbf{I}_0$}};
\draw (564.23,211.56) node [anchor=north west][inner sep=0.75pt]  [xscale=0.8,yscale=0.8] [align=left] {{\LARGE $\displaystyle M$}};
\draw (530,38.29) node [anchor=north west][inner sep=0.75pt]  [xscale=0.8,yscale=0.8] [align=left] {{\LARGE $\displaystyle T_{\mathbf{I}_0} M$}};
\draw (474.47,111.16) node [anchor=north west][inner sep=0.75pt]  [xscale=0.8,yscale=0.8] [align=left] {{\Large $\displaystyle X$}};
\draw (380,251.04) node [anchor=north west][inner sep=0.75pt]  [xscale=0.8,yscale=0.8] [align=left] {{\Large $\displaystyle \mathbf{I}_0e^{tX}$}};

\end{tikzpicture}

%% file: Simulations.tex
We first compare the performance of the Exponential Map and EM estimators using synthetic data, and 
then we show an application of the Exponential Map estimator to an example of real interferometric data. 
All computations 
and simulations were carried out in an Intel i5 7th Gen processor standard PC with 16GB of RAM memory.

\begin{figure}[h!]
    \centering
    \includegraphics[width=8.5cm, height=6.5cm]{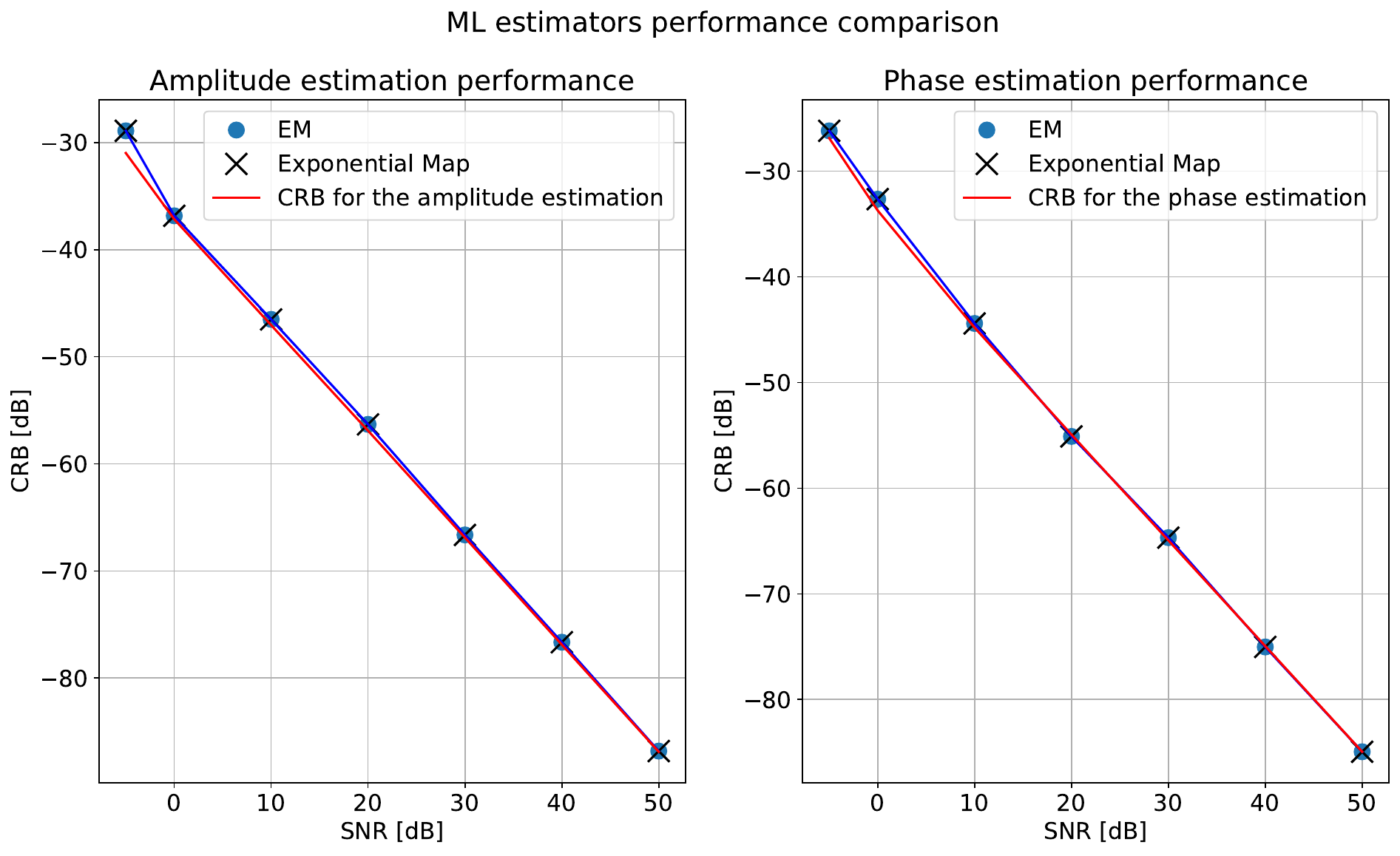}
    \caption{Comparison of Exponential Map (black crosses) and EM (blue dots) estimators with the CRB of 
    the estimation on the amplitude (left panel) and phase components (right panel) of the perturbation 
    matrix, when considering DIE only perturbations.}
    \label{fig:CRB}
\end{figure}

We assumed an interferometer with only three elements and synthesized realizations of a signal 
coming from one source located at the center of the map, with the
SNR ranging from $-5$ dB to $50$ dB ($10$ dB steps except for the first step of $5$ dB). In Fig. 
\ref{fig:CRB}, we show the variances obtained empirically togheter with the CRB. 
As expected, both estimators have the same performance and attain the CRB as they are 
assymptotically 
optimal. 

Regarding the computational time, the average 
time for the $0$ dB case was approximately $0.15$ seconds for the Exponential Map, and 
$5.8$ seconds for the EM. For the $50$ 
dB case, it took an average time of $0.12$ seconds for the Exponential Map and $5$ seconds 
approximately for the 
EM. We used a gain step $g = 0.1$ for both estimators, which was previously tested to be the 
optimal gain to perform the minimizations of Eq. (\ref{eq:NormaDIE2}) and Eq. (\ref{eq:EMmin}) in the 
minimum time. It can be seen that the Exponential Map estimator has the same performance of the EM but 
it is much faster (approximately $40$ times faster).

\begin{figure}[h!]
    \centering
    \includegraphics[width=8.5cm, height=5.5cm]{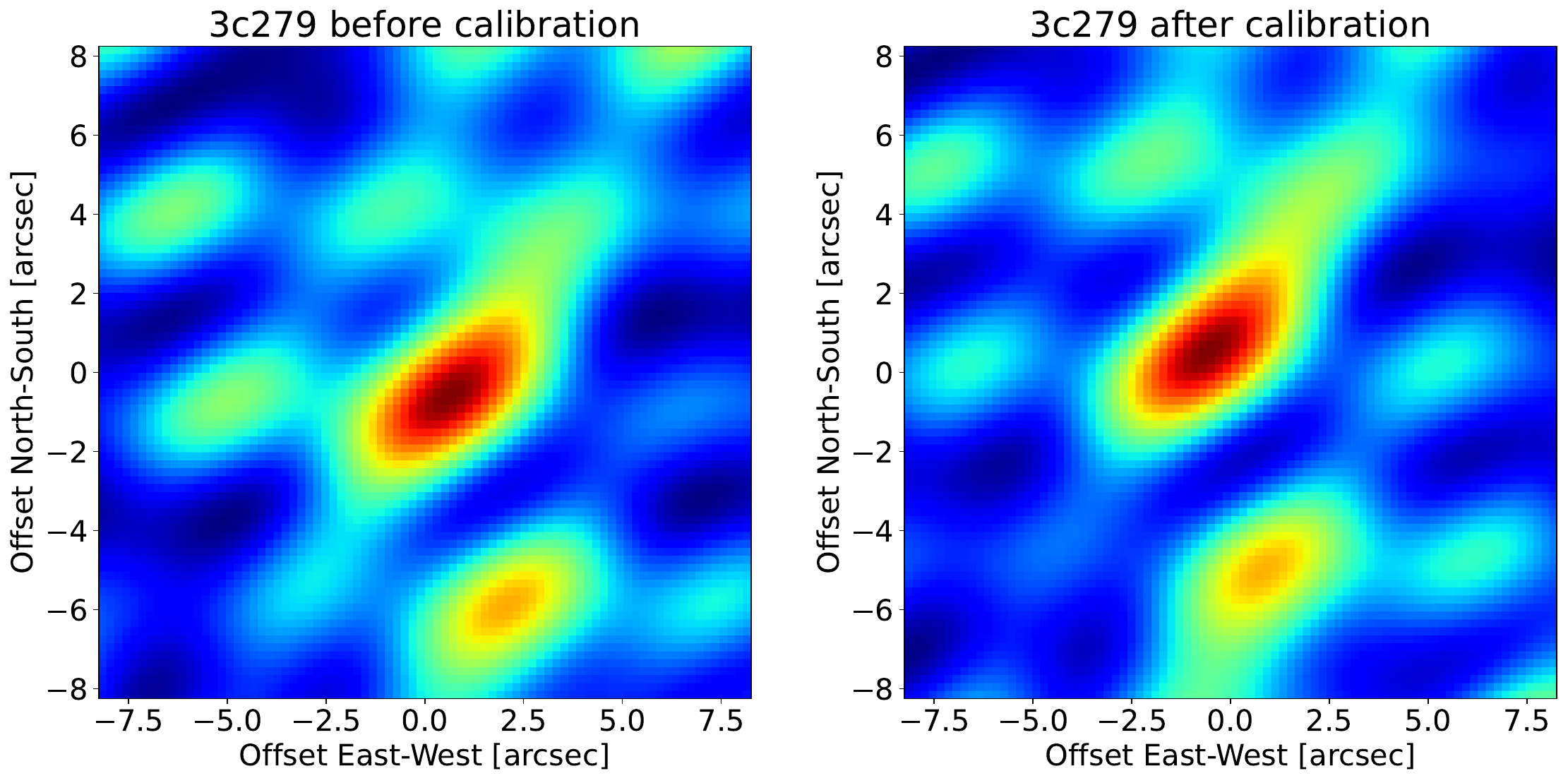}
    \caption{3c279 image before calibration (left panel) and after calibration (right panel).}
    \label{fig:3c279}
\end{figure}

\begin{figure}[h!]
    \centering
    \includegraphics[width=8.5cm, height=6.5cm]{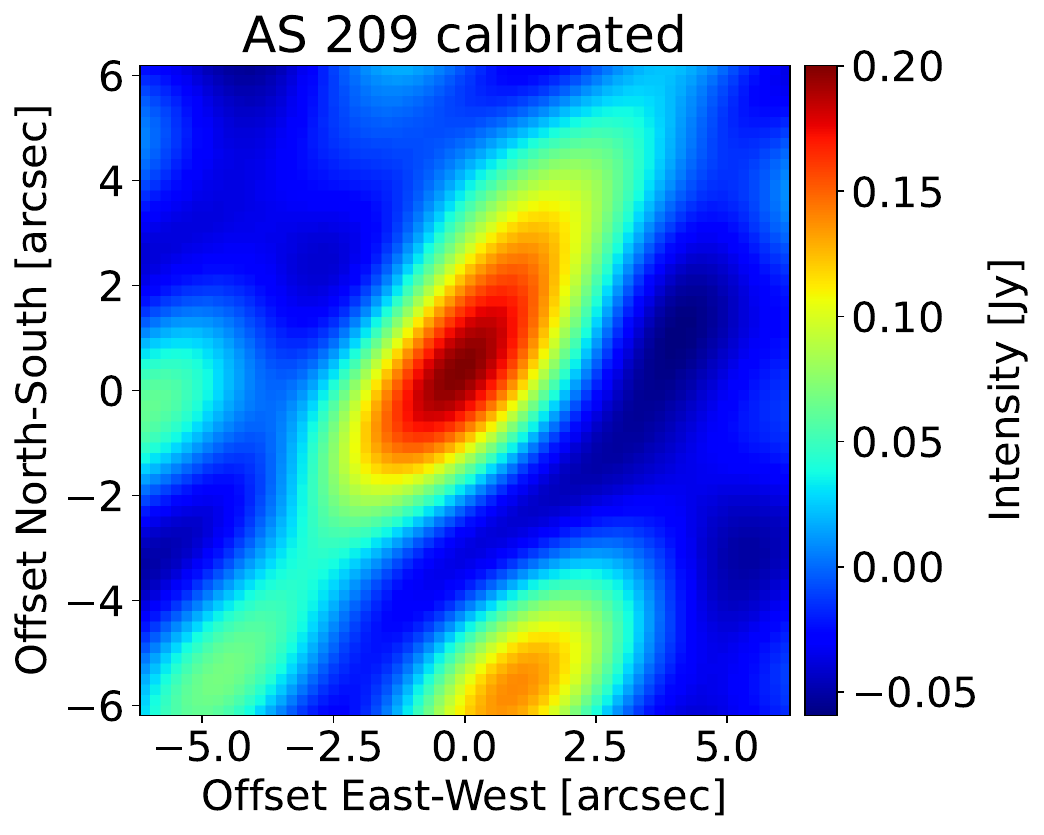}
    \caption{Image of AS 209 calibrated with the perturbation matrix obtained by 
    calibrating with 3c279.}
    \label{fig:AS209}
\end{figure}

Finally, we tested our estimator with real data. To this purpose, we used 
an observation of the protoplonanetary disk AS 209 made with the Submillimeter Array (SMA)
\footnote{This observation was carried out in the frame of the ``2023 Submillimeter Array Interferometry 
School'', where L. Cabral participated.} at 
a center 
frequency of $225$ GHz. For the computations we used only one spectral window and averaged both 
polarizations to 
work with the intensity. For imaging, we first get the visibilities using the CASA \cite{visread} 
software and then we construct the correlation matrix $\mathbf{R}$ and average over all samples of 
the spectral window. The resolution applied in the imaging process was $0.2$ arcsec and, due to 
the array resolution being larger than the source angular size, in all images we see a copy of the SMA 
Point Spread Function (PSF). Second, we obtained 
the perturbation matrix $\mathbf{G}$ by applying our proposed Exponential Map estimator with the 
calibrator source 3c279, which is the calibrator data present in the observation data of AS 209 that we 
used in this example. We then 
corrected the data of AS 209 using the estimation of $\mathbf{G}$. In Fig. \ref{fig:3c279}, it can be 
seen in the left panel the 
image of 3c279 without the calibration, and in the right panel the image of 3c279 after the calibration. 
It can be noted that the position of the source in the map and its shape are modified after the 
correction. 
After the calibration we multiplied $\mathbf{G}$ by a correction factor to match the known intensity 
of 3c279 (which was approximately $0.13$). 
Then, we used this matrix to image AS 209 as it can be seen in Fig. \ref{fig:AS209}, where the 
peak intensity 
is about $0.2$ Jy, which corresponds to previous observations at the continuum emission of this source 
made at $284$ GHz \cite{Qi}. Thus, we confirmed that the estimator worked properly for this example, which 
it also serves as an example of usage of the proposed estimator.

